\documentclass[aps,
prd,
twocolumn,
superscriptaddress,
preprintnumbers,
nofootinbib,
amsmath,amssymb,floatfix]{revtex4-2}
\usepackage{graphicx}
\usepackage{xcolor}

\usepackage[normalem]{ulem}
\usepackage{booktabs}
\usepackage[colorlinks=true,allcolors=blue]{hyperref}

\begin{document}
\preprint{INR-TH-2026-019}
\title{Cross-experiment tests of the atmospheric-muon excess}
\author{G.~I.~Rubtsov}
\affiliation{Institute for Nuclear Research of the Russian Academy of Sciences, Moscow 117312, Russia}
\author{S.~V.~Troitsky}
\email{st@inr.ac.ru}
\affiliation{Institute for Nuclear Research of the Russian Academy of Sciences, Moscow 117312, Russia}
\affiliation{Physics Department, Lomonosov Moscow State University, Moscow 119991, Russia}

\begin{abstract}
High-energy cosmic rays produce atmospheric muons through cascades of hadronic interactions, yet simulations often underestimate the muon content of extensive air showers. Whether the same problem affects energetic atmospheric muons, which probe a different part of the cascade, is less clear. Here we compare published atmospheric-muon measurements from four experiments with reference predictions based on a shared choice of hadronic interaction model and primary cosmic-ray flux. We find that the measurements predominantly exceed the central predictions across different experimental techniques,
from surface measurements to deep detectors.
This recurring excess is more consistent across the measurements than any inferred dependence on energy or zenith angle. These atmospheric muons preferentially originate from energetic mesons produced early in the cascade, whereas the total muon content of extensive air showers also reflects interactions at later stages. A common explanation of both excesses must therefore increase the effective supply of energetic muon parents, rather than merely increasing the multiplicity of soft particles. This provides a way to distinguish between competing explanations of the cosmic-ray muon puzzle.
\end{abstract}
\maketitle
\raggedbottom
\section{Introduction}
\label{sec:intro}
The muon content of extensive air showers (EAS) has long challenged hadronic-interaction models, with several experiments reporting more muons than predicted \cite{SUGAR-spark1,SUGAR_INR,HiRes-MIA,NEVOD-DECOR-2010,NEVOD-DECOR,old-Yakutsk,Auger-mu,AMIGA-mu,TA-mu}. The experimental picture is not uniform, however: some measurements agree with particular model and composition assumptions \cite{EAS-MSU_mu,new-Yakutsk,new-IceTop}, while others reveal discrepancies in the zenith-angle evolution of the muon content \cite{KASCADE-Grande}. A cross-experiment comparison by the Working Group on Hadronic Interactions and Shower Physics (WHISP) identified an increasing muon excess above primary energies of about 10~PeV \cite{Dembinski2019}. Possible explanations involve uncertainties in hadronic particle production and in the fraction of energy retained in the hadronic cascade \cite{Albrecht2022}.

Inclusive atmospheric muons probe a complementary part of the cascade. Because the primary spectrum is steep, a high-energy muon preferentially comes from an energetic parent meson produced in an early cascade generation. Contributions requiring a much more energetic primary followed by substantial energy loss are suppressed. Atmospheric-muon spectra are therefore sensitive to forward production of charged mesons and to the energy carried onward by leading hadrons \cite{Dedenko:2014lva,Dedenko:2019umk,Fedynitch2019,FedynitchHuber2022}. Recent deep-detector measurements \cite{KM3NeT2024,Zhang2025} provide further tests of this part of the shower.

Here we compare atmospheric-muon measurements by BESS \cite{Haino2004}, L3+C \cite{L3_2004}, KM3NeT \cite{KM3NeT2024}, and Jinping \cite{Zhang2025} with published reference predictions. We use published calculations based on the hadronic interaction model SIBYLL~2.3d \cite{Riehn2020} and the 2019 version of the Global Spline Fit (GSF19) of the primary cosmic-ray flux and composition \cite{Dembinski2017}. This combination was selected because it provides the broadest set of publicly available predictions for the experiments considered, rather than because of any assumed preference for these particular models. The calculations retain experiment-specific treatments of low-energy interactions, muon propagation, and detector response. Following the cross-experiment approach of WHISP, we ask which features recur across different experimental techniques: an overall excess, an energy dependence, or a zenith-angle dependence. We use \emph{muon excess} to mean that the measurement exceeds the central prediction. The comparison preserves the published absolute normalizations and connects the observed discrepancies to the parts of the hadronic cascade that these measurements probe.

The rest of the paper is organized as follows. In Sec.~\ref{sec:data}, we describe the experimental data sets, reference predictions, and construction of the comparison samples. Section~\ref{sec:results} presents the measured-to-predicted muon ratios and examines their energy and angular dependences. In Sec.~\ref{sec:physics}, we discuss the implications for hadronic particle production and possible explanations of the EAS muon puzzle. We summarize our conclusions in Sec.~\ref{sec:concl}. Technical details of the data treatment, uncertainty analysis, and cascade interpretation are given in the Appendices.

\section{Measurements and reference predictions}
\label{sec:data}
Experiments may report different observables proportional to the normalization of the atmospheric muon flux: differential surface fluxes, reconstructed event rates in deep detectors, and integral fluxes reconstructed from underground measurements. To compare these observables on a common relative scale, we divide each measured quantity by the corresponding reference prediction,
\begin{equation}
 R_{\mu,j}=\frac{\mathcal O_j^{\rm data}}{\mathcal O_j^{\rm ref}}.
 \label{eq:ratio}
\end{equation}
Here $j$ labels a measurement bin, $\mathcal O_j$ is its observable, and the superscripts ``data'' and ``ref'' denote the measurement and reference calculation, respectively. Thus $R_\mu=1$ corresponds to agreement with the central prediction. Table~\ref{tab:data} summarizes the measurements and their use in our comparison. 
These ratios allow us to compare the measurements relative to their respective predictions without treating the underlying observables as equivalent.

\begin{table*}[t]
\caption{Measurements included in the comparison.
Each observable is compared with its corresponding
published reference prediction.}
\label{tab:data}
\small
\setlength{\tabcolsep}{6pt}
\renewcommand{\arraystretch}{1.15}
\begin{tabular}{@{}
l
p{0.33\textwidth}
p{0.45\textwidth}
@{}}
\toprule
Experiment & Measured observable & Use in this work \\
\midrule

BESS \cite{Haino2004}
& Differential surface muon flux
& Momentum dependence;
  $p\geq20$~GeV/$c$ for the baseline fits \\

L3+C \cite{L3_2004}
& Differential surface muon flux
  in eight zenith-angle intervals
& Momentum dependence from averaged data;
  angular dependence from individual measurements \\

KM3NeT \cite{KM3NeT2024}
& Reconstructed muon event rates
  in ORCA6 and ARCA6
& Event-rate ratios in angular groups
  with contributing surface-muon energy ranges \\

Jinping \cite{Zhang2025}
& Reconstructed integral surface-muon flux
& Integral ratios at the corresponding
  surface-muon energy thresholds \\

\bottomrule
\end{tabular}
\end{table*}

The published reference calculations share SIBYLL~2.3d and GSF19 as their hadronic-interaction and primary-flux baselines, respectively. They differ, however, in the treatment of low-energy interactions, muon propagation, and detector response. We retain these experiment-specific configurations rather than attempting to construct a new uniform simulation. The relevant details are described below, with further technical information in Appendix~A.

For BESS \cite{Haino2004} and the cosmic-ray extension of the L3 detector at LEP (L3+C) \cite{L3_2004}, we reconstruct the surface-flux curves in Fig.~10 of Ref.~\cite{FedynitchHuber2022}, calculated with the Matrix Cascade Equations (MCEq) solver \cite{Fedynitch2015}. We convert the reference from kinetic energy to momentum to match the measured spectra, and use the appropriate angular curve for each sample. BESS changes its angular selection at 20~GeV/$c$; points above this boundary define our baseline energy comparison, while the lower-energy points test its sensitivity to the selected range. For L3+C, averaging nearby momenta across zenith intervals gives a compact energy comparison; the angular tests retain the original measurements. The published Bartol hadronic-uncertainty band \cite{Barr2006} supplies the upper and lower reference variations. Its unknown correlations are tested separately in Appendix~\ref{app:statistics}.

For KM3NeT, we combine the published fine angular bins into broader groups using the predicted event rates as weights. 
The accompanying surface-muon energy distributions show which energies contribute to each group. 

Jinping provides integral muon fluxes above surface-energy thresholds, reconstructed from underground measurements for different angular and overburden selections \cite{Zhang2025}. Details of the uncertainty treatment are given separately for each experiment in Appendix~\ref{app:data}.

All ratios retain the published absolute flux or event-rate normalization. WHISP cross-calibrated shower-energy scales by matching the measured primary cosmic-ray spectra \cite{Dembinski2019}; atmospheric-muon fluxes have no equivalent external calibration in this comparison. Shared primary-flux and hadronic uncertainties, together with incomplete covariance information, make separate comparisons more informative here than a combined significance.

\section{Results}
\label{sec:results}
\begin{figure*}[!tbp]
\centering
\includegraphics[width=0.85\linewidth]{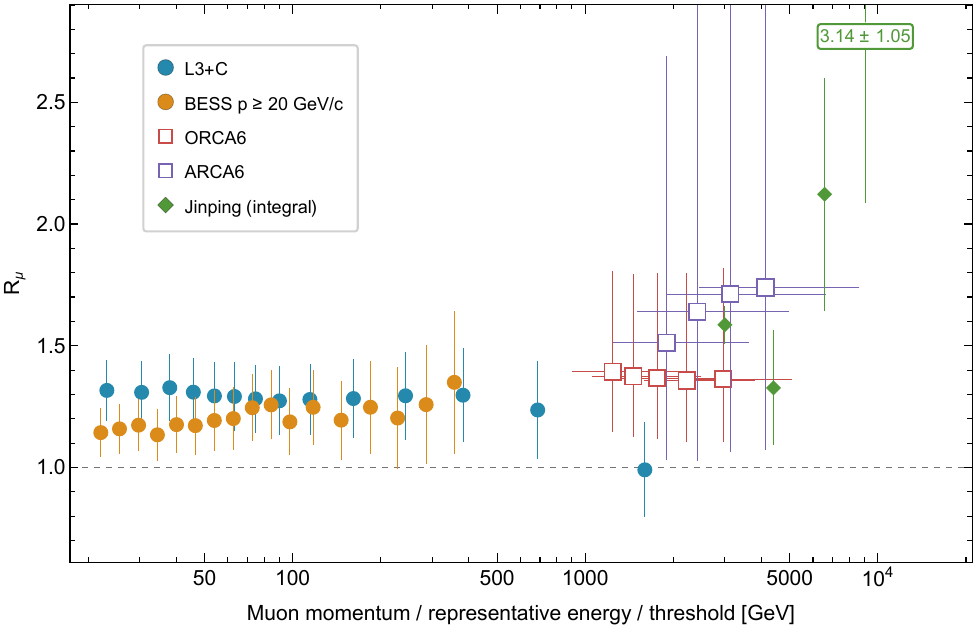}
\caption{Ratios of atmospheric-muon measurements to central SIBYLL~2.3d+GSF19-based predictions. The different observables are compared by the size of their discrepancies. For the surface spectra, vertical error bars combine experimental errors, the retained L3+C angular scatter, and the model uncertainty band. For KM3NeT, they combine statistical errors with the published reference variations propagated coherently within each angular group. KM3NeT horizontal bars span the central 68\% of the contributing sea-level muon-energy distributions. Jinping measurements are placed at their surface-energy thresholds $E_{\rm th}$ and represent fluxes reconstructed from underground measurements and integrated above these thresholds, and their error bars. The quoted value indicates the rightmost Jinping point which is above the displayed vertical range.}
\label{fig:energy}
\end{figure*}

Figure~\ref{fig:energy} compares the measured-to-predicted muon ratios across four experiments. Despite the differences in their observables and experimental conditions, the measurements show a recurring excess over the central reference predictions. The differential surface fluxes measured by L3+C give ratios mostly around 1.2--1.3, while BESS also lies predominantly above unity. For the deep-sea KM3NeT detectors, the ratios range from 1.36 to 1.39 for ORCA6 and from 1.51 to 1.74 for ARCA6. The integral fluxes reconstructed by Jinping from underground measurements show a similar excess, although their uncertainties increase at the highest thresholds. The horizontal coordinates in Fig.~\ref{fig:energy} represent different energy measures for the different observables, thus the figure only illustrates the recurrence and magnitude of the discrepancies. A related surface-flux tension was found in the underground-intensity study of Ref.~\cite{Fedynitch2022Underground}.

The magnitude of the observed excess depends on the choice of reference model. For Jinping, the measured underground muon flux exceeds the predictions of SIBYLL~2.3d, EPOS-LHC \cite{Pierog2015}, and QGSJET-II-04 \cite{Ostapchenko2011} by factors of $1.44\pm0.07$, $1.38\pm0.07$, and $1.51\pm0.08$, respectively \cite{Zhang2025}. Accounting for the quoted uncertainties in the primary cosmic-ray spectrum and hadronic interactions leaves a discrepancy with a significance exceeding $2\sigma$. The same study finds that more recent models, including EPOS-LHC-R \cite{Pierog2023}, QGSJET-III \cite{Ostapchenko2024}, and SIBYLL$^\star$ \cite{Riehn2024Star}, produce only modest changes in the predicted yield of muons above 3~TeV. A more systematic assessment of the model dependence, including the impact of recent updates to the primary cosmic-ray flux parametrization \cite{GSF2026}, will require consistently recalculated predictions for the individual experiments.

We next examine whether the observed excess exhibits systematic  dependences on muon momentum or zenith angle.

\subsection{Energy dependence and relative flux levels}
\label{sec:energy}
The surface measurements (BESS and L3+C) show a muon excess, but do not establish whether it increases with momentum. To quantify a possible energy dependence, we fit the differential surface-muon ratios with
\begin{equation}
 R_\mu(p)=b+a\log_{10}(p/p_0),
 \qquad p_0=1~{\rm GeV}/c.
 \label{eq:fit}
\end{equation}
Using experimental uncertainties alone and a common constant term $b$ for both experiments, we obtain $a=0.080\pm0.020$ ($\chi^2/{\rm dof}=56.5/29$), suggesting an increase of the excess with momentum. However, this result is largely driven by the difference between the absolute flux levels of the two experiments. At 50~GeV/$c$, their independently fitted levels differ by about 9\%. Since BESS contributes mostly at lower momenta and L3+C extends to higher momenta, this offset can mimic a rising ratio in the combined data. Allowing separate constant terms for the two experiments reduces the slope to $a=0.002\pm0.024$, consistent with no momentum dependence.

The significance of the apparent rise is also sensitive to the treatment of model uncertainties. Treating the published model uncertainty independently at each point gives $a=0.012\pm0.065$ with a common constant term and $a=-0.019\pm0.067$ with separate constant terms. The results also depend on the assumed correlations of the model uncertainties, which are not provided with the published predictions. Thus, while the surface measurements  consistently exceed the central predictions, they do not provide robust evidence for a hardening of the muon spectrum relative to the reference model. Establishing such a hardening would require better control of the relative experimental normalizations and the correlations of the theoretical uncertainties. The fit results and uncertainty-correlation tests are presented in Appendix~\ref{app:statistics} (Table~\ref{tab:energyfits}), with additional sensitivity tests in Appendix~\ref{app:sensitivity}.

\subsection{Zenith-angle dependence}
\label{sec:angular}
The L3+C measurements do not establish a smooth zenith-angle dependence of the muon excess, but reveal a localized deviation near the vertical direction. Eleven momentum groups, spanning approximately 46--1600~GeV/$c$, contain measurements in all eight zenith-angle intervals. Within each group, the ratios are compatible with a constant when diagonal experimental uncertainties are used: the corresponding $\chi^2$ values range from 2.0 to 5.4 for seven degrees of freedom.

\begin{figure}[!tbp]
\centering
\includegraphics[width=\linewidth]{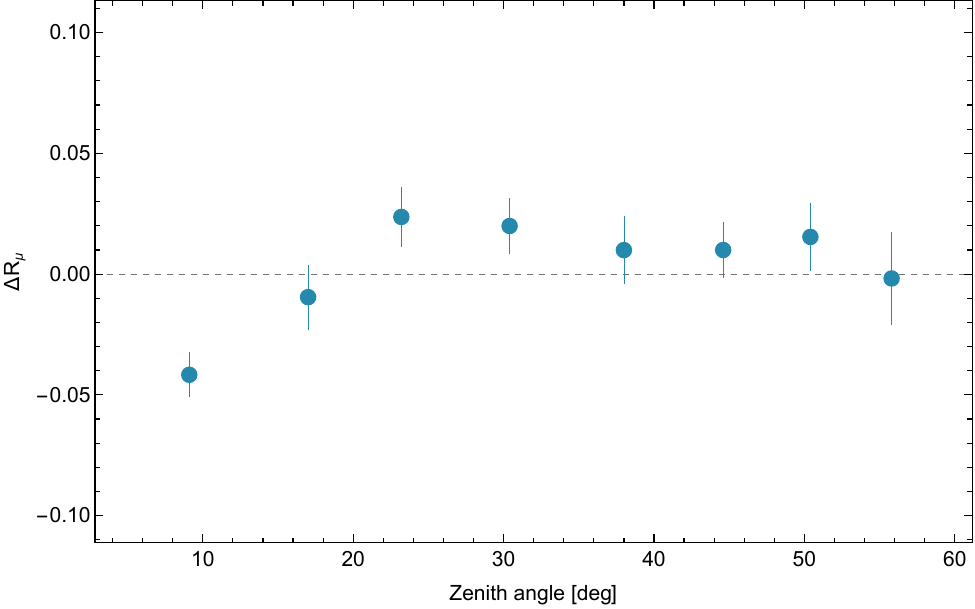}
\caption{Angular residuals $\Delta R_\mu$ of the L3+C
measured-to-predicted muon ratios, averaged over
eleven momentum groups. In each group, the
best-fitting constant $\bar R_{\mu,p}$ is subtracted
before averaging. Error bars are obtained by propagating the
diagonal experimental uncertainties through the
subtraction, including their covariance with the fitted
group constants.}
\label{fig:angles}
\end{figure}

To identify a possible angular pattern common to different momenta, we subtract the best-fitting constant $\bar R_{\mu,p}$ in each momentum group and average the residuals $\Delta R_\mu=R_\mu-\bar R_{\mu,p}$ over the eleven groups. As shown in Fig.~\ref{fig:angles}, the most vertical interval lies systematically below the others.
Fitting the original 88 ratios with eleven independent constant levels and one common angular slope gives
\begin{equation}
 \frac{dR_\mu}{d\cos\theta}=-0.090\pm0.035
 \label{eq:angular_diag}
\end{equation}
under the same diagonal-error approximation. However, omitting the most vertical interval, $0.975<\cos\theta<1$, changes the slope to $+0.002\pm0.043$, consistent with zero. The apparent angular trend is therefore driven almost entirely by this single interval. Its statistical interpretation is complicated by the correlations of the experimental and
theoretical uncertainties. Using the published experimental covariance with the central reference prediction gives a poor global fit, while including model uncertainties leads to results that depend strongly on their assumed correlations.
We therefore regard the vertical-bin deviation as the main empirical feature, rather than evidence for a broader angular distortion.  The individual angular distributions and uncertainty tests are presented in Appendix~\ref{app:statistics}.

\section{Physical interpretation}
\label{sec:physics}
The atmospheric-muon excess adds a physically different test of hadronic interaction models to the EAS muon puzzle. Most muons counted in a high-energy air shower are relatively soft and are produced after many interactions. Their number is sensitive to the energy retained in the hadronic cascade at each generation. By contrast, an energetic atmospheric muon observed inclusively is preferentially supplied by an energetic parent in the early stages of a shower. Producing the same muon after substantial energy degradation requires a rarer, higher-energy primary. This selection is the reason that atmospheric-muon spectra can help distinguish microscopic explanations of the excess.

Because the primary cosmic-ray spectrum is steep, the contribution of secondary mesons to the high-energy muon flux depends strongly on the fraction of the projectile energy they carry. In the scaling approximation, this dependence is described by the spectrum-weighted production moment \cite{Gaisser2013,Fedynitch2015},
\begin{equation}
 Z_{Nh}(E_N)=\int_0^1 dx\,x^{\gamma-1}
 \frac{dn_{N\to h}(E_N,x)}{dx},
 \label{eq:z}
\end{equation}
where $\phi_N\propto E_N^{-\gamma}$ is the differential primary-nucleon flux per unit nucleon energy $E_N$, with units of ${\rm GeV}^{-1}\,{\rm m}^{-2}\,{\rm s}^{-1}\,{\rm sr}^{-1}$, $h$ denotes a secondary hadron, and $x=E_h/E_N$. For $\gamma\simeq2.7$, the weight $x^{1.7}$ strongly suppresses contributions from secondaries carrying a small fraction of the projectile energy. Thus, the production of energetic muons is sensitive not simply to the multiplicity of secondary mesons, but to their energy distribution, particularly in the forward region.

This distinction constrains hadronic explanations that seek to account for both the atmospheric-muon excess and the EAS muon puzzle. Increasing the production of soft particles at late stages of the cascade may raise the total EAS muon number without appreciably affecting the flux of energetic atmospheric muons. A common explanation must also increase the effective supply of energetic parent mesons. This may involve changes in forward charged-meson production or in the energy carried by leading baryons into subsequent interactions~\cite{FedynitchHuber2022}. Mechanisms such as enhanced $\rho^0$ production at the expense of neutral pions or increased baryon--antibaryon production can retain more energy in the hadronic cascade, but their effect on energetic muons depends on how this energy is distributed among secondary particles. Indeed, studies with phenomenologically modified versions of SIBYLL~2.3d, collectively known as SIBYLL$^\star$, show that substantial increases in EAS muon numbers can coexist with only modest changes in inclusive atmospheric-muon spectra \cite{Riehn2024Star}.

Our analysis does not establish a significant momentum dependence of the atmospheric-muon excess (Sec.~\ref{sec:energy}). Such a dependence, if confirmed, could provide an additional test of hadronic-production mechanisms. In particular, a rising ratio $R_\mu(p)$ would imply a harder muon spectrum than the reference prediction and could reflect an energy-dependent change in forward meson production. However, changes in the primary nucleon spectrum could produce a similar effect. The primary composition  matters, as it determines the supply of nucleons at a given energy per nucleon (see Appendix~\ref{app:physics}). Distinguishing these possibilities would require better control of both the primary cosmic-ray flux and the experimental and theoretical uncertainties.

The L3+C data show a lower muon ratio in the most vertical zenith-angle interval, but provide no robust evidence for a systematic dependence on zenith angle (Sec.~\ref{sec:angular}). The angular distribution is sensitive to the competition between meson decay and reinteraction. Because pions and kaons have different critical energies, approximately 115 and 850~GeV, changes in their relative contributions can modify the angular dependence of the muon flux. A comparable enhancement of both contributions could instead increase the flux without substantially changing its angular shape. At sufficiently high energies, the conventional muon flux approaches a $1/\cos\theta$ dependence, whereas a prompt component is approximately isotropic. An enhanced prompt contribution would therefore tend to produce a larger fractional excess toward the vertical, opposite to the observed L3+C feature. However, this qualitative argument does not provide a constraint on prompt muon production, which would require a joint spectral and angular analysis with properly correlated uncertainties. The relevant cascade expressions are given in Appendix~\ref{app:physics}.

The excess in the absolute atmospheric-muon flux does not by itself distinguish enhanced hadronic production from a higher primary nucleon flux. Complementary observables offer ways to disentangle these effects. The muon charge ratio is sensitive to charge-asymmetric meson production, while largely cancelling a common flux normalization \cite{Fedynitch2019}. Atmospheric neutrinos probe different relative contributions from pion and kaon decays than muons, so a joint comparison of their spectra can help distinguish enhanced kaon production from changes in the primary flux \cite{Riehn2024Star}. Seasonal variations provide an additional probe of the competition between meson decay and reinteraction \cite{GaisserSeasonal2021,Gaisser2021,NOvA2026}. Together, these measurements can constrain whether hadronic mechanisms proposed to resolve the EAS muon puzzle also enhance the production of energetic muon parents.

\section{Conclusions}
\label{sec:concl}
The atmospheric-muon measurements considered here show an excess relative to their central reference predictions in all four experiments. The recurrence of this excess across different observables, experimental techniques, and overburdens is the main empirical result of our comparison.

The available data do not establish a robust momentum dependence of the excess. The apparent rise in the combined BESS and L3+C data is largely driven by a difference in their absolute flux levels, which can mimic an energy dependence because the experiments cover different momentum ranges. When their levels are fitted independently, the slope is consistent with zero. The L3+C measurements show a lower ratio in the most vertical zenith-angle interval, but no robust evidence for a systematic dependence on zenith angle. A more quantitative interpretation requires better knowledge of the experimental and theoretical uncertainty correlations.

These results provide a complementary test of hadronic explanations of the EAS muon puzzle. Energetic atmospheric muons preferentially probe the production of energetic parent mesons in the early stages of the cascade, whereas the total EAS muon number is also sensitive to subsequent interactions. Complementary measurements, including muon charge ratios, atmospheric-neutrino spectra, and seasonal variations, can help distinguish changes in hadronic production from uncertainties in the primary cosmic-ray flux. A hadronic mechanism invoked to explain both excesses must therefore increase the effective supply of energetic meson parents, rather than merely increase the final muon number in EAS.

\section*{Acknowledgments}
We dedicate this paper to the memory of our senior colleague, L.G.~Dedenko, who pioneered atmospheric-muon tests of the muon excess. The authors acknowledge the use of publicly available large language models for improving the language and style of this paper.
This work was supported by the Russian Science Foundation, grant 22-12-00333.
\section*{Data Availability Statement}
The analysis is based on experimental measurements and theoretical predictions published in the references cited in the text. Some numerical inputs were reconstructed from published figures. These reconstructed data and the analysis scripts are available from the corresponding author upon reasonable request, subject to applicable third-party rights.
\appendix
\section{Construction of the comparison samples}
\label{app:data}
\subsection{Surface-muon flux predictions and momentum conversion}
For BESS and L3+C, we reconstruct the predicted surface-muon spectra from the dashed SIBYLL~2.3d curves in Fig.~10 of Ref.~\cite{FedynitchHuber2022}. These calculations use GSF19 primary fluxes and DPMJET-III for projectile energies below 80~GeV, while SIBYLL 2.3d is used above this threshold. The associated uncertainty bands are based on the Barr-parameter prescription \cite{Barr2006}. We extract the central curves and the boundaries of their uncertainty bands separately from the vector graphics of the published PDF. Interpolation is restricted to the displayed kinetic-energy range $3 \leq T\leq3\times10^4$~GeV.

The theoretical curves give the differential muon flux per unit kinetic energy, $\phi_T(T)$, whereas the experiments report the differential flux per unit momentum, $\phi_p(p)$. We convert the published quantity $T^3\phi_T$ to $p^3\phi_p$ using the standard relativistic relation between kinetic energy and momentum. Our default procedure interpolates $T^3\phi_T$ linearly in $\log T$ along the extracted curves. Log--log linear interpolation and shape-preserving cubic interpolation are used to test the sensitivity to this choice.

\subsection{BESS}
\label{app:bess}
We obtain the total muon flux by summing the charge-separated measurements of Ref.~\cite{Haino2004}. Statistical uncertainties are combined in quadrature, while systematic uncertainties common to the two charges are treated coherently. The published acceptance and spectral-deformation uncertainties are also included. The BESS angular selection changes at 20~GeV/$c$: $\cos\theta>0.98$ below this momentum and $\cos\theta>0.90$ above it. For the baseline sample, $p\geq20$~GeV/$c$, we use the calculated flux at $17^\circ$, close to the mean direction of the wider acceptance. Averaging the reference curves at $9.1^\circ$, $17^\circ$, and $23.2^\circ$ over the corresponding angular intervals, assuming uniform acceptance in solid angle, changes the individual ratios by less than 0.5\%.

The lower-momentum measurements are used only for sensitivity tests (Appendix~\ref{app:sensitivity}). For $5\leq p<20$~GeV/$c$, the $9.1^\circ$ reference is the only available calculated direction within the BESS acceptance, $\theta<11.48^\circ$. To estimate the effect of using this fixed-angle approximation, we extrapolate the angular dependence of the predicted flux toward the vertical. We interpolate its logarithm as a linear function of $\cos\theta$ using the $9.1^\circ$ and $17^\circ$ curves, or as a quadratic function using these curves and the one at $23.2^\circ$. Averaging the resulting fluxes uniformly over $0.98<\cos\theta<1$ lowers the measured-to-predicted ratios by  less than 0.4\%.
We note, however, that these estimates depend on the assumed angular extrapolation and do not constitute rigorous bounds on the reference uncertainty. The 20~GeV/$c$ boundary is determined by the experimental angular selection and is unrelated to the 80~GeV projectile-energy threshold between the low- and high-energy hadronic models.

\subsection{L3+C}
The L3+C measurements \cite{L3_2004} cover eight zenith-angle intervals, for which we use the corresponding calculated muon spectra. The reported normalization uncertainties include 0.7\% from live time and trigger and scintillator efficiencies, and (1.7--3.7)\% from detector acceptance, depending on zenith angle. An independent check using $Z\to\mu^+\mu^-$ events constrains the absolute normalization to within 5.2\% at 68\% CL. We treat this result as a validation of the normalization rather than an additional source of uncertainty.

For the momentum-dependence analysis, we combine measurements from different zenith-angle intervals whose momenta differ by less than 1.5\%. The measured-to-predicted ratios are averaged using inverse-variance weights based on the diagonal experimental uncertainties. Statistical errors are propagated in quadrature, whereas systematic errors and shifts due to the uncertainty in the predicted flux are combined coherently. To account for the dispersion among the angular measurements, we add a scatter contribution to the variance,
\begin{equation}
 \sigma_{\rm scatter}^2=
 \frac{\sum_i w_i(R_i-\bar R)^2}{N_{\rm eff}},
 \qquad
 N_{\rm eff}=\frac{1}{\sum_i w_i^2},
 \label{eq:l3scatter}
\end{equation}
where the weights are normalized to $\sum_i w_i=1$ and $\bar R=\sum_i w_iR_i$. The representative momentum of each group is the geometric mean of its measured momenta. The resulting points are used in the energy fits with diagonal uncertainties.

For the zenith-angle analysis, we retain the individual measurements rather than their momentum-group averages. The published L3+C uncertainties include systematic correlations between zenith-angle intervals and across momentum bins, as well as statistical correlations between neighboring momentum bins \cite{L3_2004}. We reconstruct the experimental covariance matrix using the correlation coefficients reported in Tables~1--9 of that reference. For momentum index $i$ and angular index $j$, the nonzero off-diagonal contributions are
\begin{align}
 C^{\rm stat}_{(i,j),(i+1,j)}
 &=\rho_i^{(j)}
   \sigma^{\rm stat}_{ij}
   \sigma^{\rm stat}_{i+1,j},
 \notag\\
 C^{\rm syst}_{(i,j),(i',j')}
 &=\rho^{\rm syst}_{jj'}
   \sigma^{\rm syst}_{ij}
   \sigma^{\rm syst}_{i'j'}.
\end{align}
The published inter-angle correlation matrix, given to limited numerical precision, has a small negative eigenvalue. We make it positive semidefinite by clipping negative eigenvalues and renormalizing the diagonal elements to unity. The largest change in an individual correlation coefficient is $1.7\times10^{-3}$. Fits using the reconstructed covariance matrix are discussed in Appendix~\ref{app:statistics}.

\subsection{KM3NeT}
For KM3NeT, we extract the measured-to-predicted event-rate ratios and variations of the reference predictions from the published vector figures \cite{KM3NeT2024}. We combine the original zenith-angle bins into broader groups, weighting their ratios by the predicted event rates $M_j$:
\begin{equation}
 R_{\mu,J}=
 \frac{\sum_{j\in J}M_jR_{\mu,j}}
      {\sum_{j\in J}M_j},
 \label{eq:rebin}
\end{equation}
where $J$ denotes a combined angular group.  Statistical uncertainties are propagated in quadrature, while each variation of the reference prediction is applied coherently across the contributing bins.

The horizontal bars in Fig.~\ref{fig:energy} indicate the central 68\% of the published sea-level muon-energy distributions, with markers at their medians. These distributions refer to individual muons in reconstructed bundles, rather than to a single energy associated with each event. The horizontal coordinates therefore indicate the characteristic muon energies contributing to the event rates, but cannot be used directly to infer a differential energy dependence. A supplementary fit using effective energies is discussed in Appendix~\ref{app:sensitivity}.

We note that the measured-to-predicted ratios, particularly for ARCA6, are sensitive to the assumed light-absorption length in seawater \cite{KM3NeT2024}.

\subsection{Jinping}
For Jinping, we compare the integral muon fluxes above surface-energy thresholds reconstructed from underground measurements using simulated muon survival probabilities (Fig.~5 of Ref.~\cite{Zhang2025}) with the corresponding SIBYLL~2.3d predictions. Different thresholds correspond to different zenith-angle and rock-overburden selections. The resulting points therefore do not represent a differential energy spectrum. We retain the integral nature of the measurements and the published error bars for the reconstructed fluxes. The reconstruction accounts for variations among hadronic models, but the source does not give a separate breakdown of uncertainties for these points. The Jinping ratios enter the overall comparison in Fig.~\ref{fig:energy}, but not the momentum-dependence fits, which use only the differential surface spectra.

Throughout the comparison, we retain the original particle content of each published prediction, including any prompt-muon contribution already present.

\section{Uncertainty and correlation tests}
\label{app:statistics}
\subsection{Surface-energy fits}
Table~\ref{tab:energyfits} summarizes four fits to the momentum dependence of the BESS and L3+C ratios, using the parametrization of Eq.~\eqref{eq:fit}. We consider either a common constant term or separate constant terms for the two experiments, with a common slope in both cases. For each choice, the fits are performed with experimental uncertainties alone or with the published model uncertainty added independently at each point.
\begin{table}[!tbp]
\caption{Fits to the momentum dependence of the BESS ($p\geq20$~GeV/$c$) and L3+C surface-muon ratios. All fits use diagonal experimental variances, including the L3+C angular scatter. ``Pointwise'' denotes the addition of model uncertainties independently at each point. separate constant terms allow different absolute levels for the two experiments.}
\label{tab:energyfits}
\footnotesize
\begin{ruledtabular}
\begin{tabular}{lccc}
Model uncertainty & Intercept & $a$ & $\chi^2/{\rm dof}$ \\
\hline
Omitted & Common & $0.080\pm0.020$ & $56.5/29$ \\
Omitted & Independent & $0.0019\pm0.0244$ & $24.2/28$ \\
Pointwise & Common & $0.012\pm0.065$ & $7.70/29$ \\
Pointwise & Independent & $-0.019\pm0.067$ & $4.22/28$ \\
\end{tabular}
\end{ruledtabular}
\end{table}

The experimental variances are constructed as
$\sigma_{i,\rm exp}^2=
\sigma_{i,\rm stat}^2+
\sigma_{i,\rm syst}^2+
\sigma_{i,\rm scatter}^2$,
where the scatter term applies to the averaged L3+C measurements. In the pointwise fits, the appropriate upper or lower displacement of the model-uncertainty band, converted to ratio units, is added in quadrature. The asymmetric displacement is selected according to the direction of the trial fitted value. Parameter uncertainties correspond to an increase of one in the profiled $\chi^2$. This procedure neglects correlations of the model uncertainties across momentum.

With separate constant terms, the fitted ratios at 50~GeV/$c$ are 1.178 for BESS and 1.284 for L3+C using experimental uncertainties alone, or 1.189 and 1.287 when the pointwise model uncertainties are included. These differences characterize the relative levels of the two data sets; they should not be interpreted as allowed calibration shifts. A fit with constrained normalization adjustments would require the experimental calibration uncertainties and their correlations, avoiding double counting of systematic errors already included in the flux measurements. Data-driven approaches such as daemonflux \cite{Yanez2023} provide a framework for incorporating calibration information.

To examine the effect of theoretical correlations, we retain the same experimental variances but represent the model-uncertainty band through Gaussian nuisance parameters. We first symmetrize the published band in logarithmic flux,
\begin{align}
 u_i&=\frac12\ln\!\left(
 \frac{\phi_i^{\rm upper}}{\phi_i^{\rm lower}}
 \right),\notag\\
 R_i^{\rm pred}
 &=\left[b+a\log_{10}(p_i/p_0)\right]
 e^{\lambda_i u_i},
 \label{eq:energy_theory_nuisance}
\end{align}
where $\phi_i^{\rm upper}$ and $\phi_i^{\rm lower}$ denote the boundaries of the predicted flux band. We compare two limiting assumptions: independent standard-normal parameters $\lambda_i$ for individual points, and a single common parameter $\lambda_i=\lambda$ for all BESS and L3+C points. The corresponding Gaussian penalties are $\sum_i\lambda_i^2$ and $\lambda^2$, respectively. The model uncertainty enters only through these parameters, without an additional contribution to the experimental variance. For averaged points, the band widths are reconstructed from the coherently averaged model variations.

\begin{table}[!tbp]
\caption{Sensitivity of the fitted momentum slope to assumptions about correlations of the model uncertainties. Both scenarios use the same diagonal experimental variances and represent the model-uncertainty band through Gaussian nuisance parameters.}
\label{tab:energy_theory_corr}
\footnotesize
\begin{ruledtabular}
\begin{tabular}{lcc}
Theory scenario & $a$ & $\chi^2/{\rm dof}$ \\
\hline
Independent modes & $0.019^{+0.070}_{-0.069}$ & $7.34/29$ \\
One common mode & $0.140^{+0.090}_{-0.077}$ & $55.89/29$ \\
\end{tabular}
\end{ruledtabular}
\end{table}
The results in Table~\ref{tab:energy_theory_corr} illustrate the sensitivity of the inferred slope to theoretical correlations. The common-mode fit gives a steeper positive slope, but describes the data substantially less well than the independent-mode fit. In the common-mode case, the best-fitting nuisance parameter is $\lambda=-0.779$, with contributions of 55.29 from the experimental residuals and 0.61 from the Gaussian penalty to the total $\chi^2$. The corresponding contributions for independent modes are 1.12 and 6.22. Since the width of the model-uncertainty band depends on momentum, a coherent variation changes both the overall level and the shape of the predicted ratio.

These scenarios are sensitivity tests rather than statistical bounds on the possible slopes. The published uncertainty band does not specify its covariance matrix, nor does it imply a particular Gaussian nuisance-parameter model. Consequently, the apparent momentum dependence
cannot be assigned a model-independent significance from these fits.

\subsection{L3+C zenith-angle fits}
\begin{figure*}[!tbp]
\centering
\includegraphics[width=0.93\linewidth]{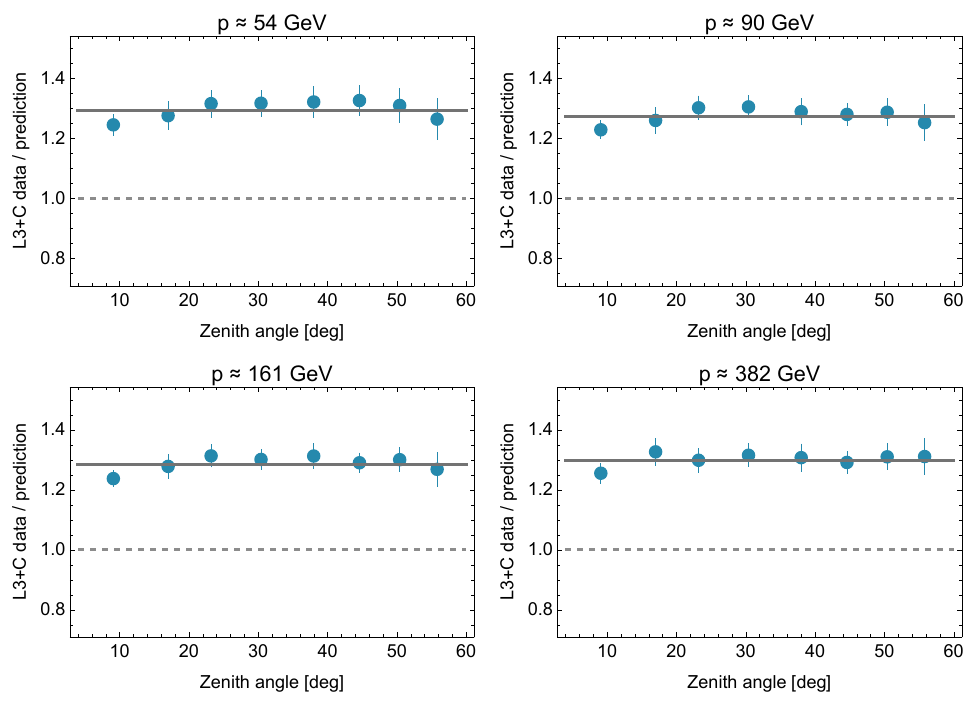}
\caption{Zenith-angle dependence of the L3+C
measured-to-predicted muon ratios for four
representative momentum groups out of eleven
used in the analysis. Error bars combine
statistical and systematic experimental
uncertainties in quadrature. Solid lines show
the best-fitting constant ratios, while
dashed lines indicate $R_\mu=1$.
The indicated momenta are representative
values for each group.}
\label{fig:matched}
\end{figure*}

The zenith-angle analysis uses 88 L3+C measurements in eleven momentum groups, each containing eight zenith-angle intervals. Figure~\ref{fig:matched} illustrates the measured-to-predicted ratios for four representative momentum groups. As described in Sec.~\ref{sec:angular}, we fit the full sample with an independent constant level for each momentum group and a common slope in $\cos\theta$. Here we examine the sensitivity of this fit to the treatment of experimental and theoretical correlations.

Using the reconstructed experimental covariance matrix alone, with the central reference prediction held fixed, gives $\chi^2/{\rm dof}=808.33/76$. This poor fit indicates that the experimental uncertainties alone cannot account for the observed deviations from the reference prediction. We therefore examine the effect of uncertainties in the predicted muon flux and its zenith-angle dependence. 

The published model-uncertainty band does not specify correlations between the measurement bins. As a sensitivity test, we take $t_i$ to be the mean magnitude of the upper and lower theoretical variations in ratio units and consider two covariance prescriptions,
\begin{equation}
 C_{\rm th}^{\rm diag}
 =\operatorname{diag}(t_i^2),\qquad
 C_{\rm th}^{\rm corr}=t\,t^{\mathsf T},
 \label{eq:theory_cov_limits}
\end{equation}
where $t$ is the vector of uncertainties $t_i$. Both matrices are added to the same experimental covariance. The diagonal prescription treats the theoretical variations as independent between points, whereas the second assumes one fully correlated variation with point-dependent amplitudes.

\begin{table}[!tbp]
\caption{Sensitivity of the fitted L3+C zenith-angle slope to assumptions about theoretical correlations. Both scenarios use the same reconstructed experimental covariance matrix.}
\label{tab:angular_theory_corr}
\footnotesize
\begin{ruledtabular}
\begin{tabular}{lcc}
Theory scenario & $dR_\mu/d\cos\theta$ & $\chi^2/{\rm dof}$ \\
\hline
Independent variations & $-0.0688\pm0.1746$ & $4.50/76$ \\
One common variation & $+0.0164\pm0.0276$ & $107.22/76$ \\
\end{tabular}
\end{ruledtabular}
\end{table}
As shown in Table~\ref{tab:angular_theory_corr}, the inferred slope depends strongly on the assumed theoretical correlations. Independent variations give a slope with a large uncertainty and a very small $\chi^2$, whereas a single correlated variation yields a smaller slope uncertainty but a substantially poorer fit. The latter uncertainty should therefore not be interpreted as a precise constraint on the zenith-angle dependence.

To test the localized deviation directly, we replace the common angular slope by an offset $\delta_{\rm vert}$ applied only to the most vertical interval, $0.975<\cos\theta<1$, while retaining a separate constant level for each momentum group. With diagonal experimental uncertainties, we obtain $\delta_{\rm vert}=-0.0529\pm0.0116$. Using the reconstructed experimental covariance alone gives $-0.0587\pm0.0027$, but with a very poor fit ($\chi^2/{\rm dof}=495.7/76$). Including independent theoretical variations gives $-0.0491\pm0.0505$ ($\chi^2/{\rm dof}=3.71/76$), whereas one common variation gives $-0.0097\pm0.0036$ ($\chi^2/{\rm dof}=100.4/76$). Thus, although the central ratios show a lower value near the vertical direction, the magnitude and statistical significance of this contrast depend strongly on the assumed theoretical correlations. A quantitative assessment of the angular dependence requires the covariance of the theoretical predictions.

\section{Additional tests of momentum dependence}
\label{app:sensitivity}
The BESS measurements below 20~GeV/$c$ are excluded from the baseline analysis because of their different angular acceptance. We use them here to test the sensitivity of the fitted slope to the momentum range. Making use of the calculated muon flux at $9.1^\circ$ for the additional points, we extend the sample down to 10 or 5~GeV/$c$ and obtain slopes of 0.118 and 0.140, respectively, for these two energy cuts (in the fit version with a common constant term and pointwise model uncertainties). Allowing separate constant terms for BESS and L3+C reduces these slopes to 0.065 and 0.097. Replacing the fixed-angle approximation by either of the two angular averages described in Appendix~\ref{app:bess} changes the common-intercept slopes by only 0.0015--0.0017 and 0.0025--0.0026 for the two momentum cuts, respectively. The fitted slope is thus considerably more sensitive to the inclusion of lower-momentum measurements than to the assumed angular dependence of the calculated flux.

For the baseline BESS sample, averaging the predicted flux over the angular acceptance, rather than using the $17^\circ$ curve, changes the common-intercept slope from 0.0118 to 0.0086 when pointwise model uncertainties are included. Using the $9.1^\circ$ or $23.2^\circ$ curves instead of $17^\circ$ changes the slope by at most 0.017, while different interpolation methods change it by less than $10^{-3}$. Fits to BESS and L3+C separately give $a=0.109\pm0.106$ and $a=-0.106\pm0.087$, respectively, with pointwise model uncertainties. Neither experiment therefore establishes a significant momentum dependence on its own.

As an additional test, we include the KM3NeT event-rate ratios in a combined fit by assigning them effective energies $E_{\rm geom}=10^{\langle\log_{10}E\rangle}$, with $E$ expressed in GeV. This gives $a=0.061^{+0.049}_{-0.048}$ for a common constant term. However, the effective energies are derived from distributions of individual muons in reconstructed bundles and do not fully represent the detector response. This fit is therefore only an approximate test and is not included in the baseline analysis of differential surface spectra. 

\section{Cascade interpretation of spectral and
zenith-angle dependences}
\label{app:physics}
The energy and zenith-angle dependences of conventional atmospheric muons can be understood from an approximate cascade expression
\cite{Gaisser2013},
\begin{equation}
\begin{split}
 \phi_\mu(E,\theta)\propto E^{-\gamma}\!\bigg[&
 \frac{A_\pi}{1+B_\pi E\cos\theta/\epsilon_\pi}\\[-2pt]
 &+\frac{A_K}{1+B_K E\cos\theta/\epsilon_K}\bigg].
\end{split}
 \label{eq:angular_physics}
\end{equation}
Here, $\phi_\mu(E,\theta)$ is the differential flux of conventional atmospheric muons per unit energy and solid angle, with $E$ and $\theta$ denoting the muon energy and zenith angle. The critical energies of pions and kaons are $\epsilon_\pi\simeq115$~GeV and $\epsilon_K\simeq850$~GeV, respectively. The coefficients $A_h$ determine the normalization of the pion and kaon contributions. In the scaling approximation, they contain the production moments $Z_{Nh}$ divided by $1-Z_{NN}$, multiplied by the corresponding meson-decay moments \cite{Fedynitch2019,FedynitchHuber2022}. Here, $Z_{NN}$ describes nucleon regeneration. The coefficients $B_h$ depend on hadronic attenuation lengths and decay kinematics and characterize the transition between meson decay and reinteraction.

At energies satisfying $B_hE\cos\theta\gg\epsilon_h$ for both meson species, the conventional flux approaches a $1/\cos\theta$ dependence. An approximately isotropic prompt component would then contribute a larger fraction of the total flux toward the vertical, as discussed in Sec.~\ref{sec:physics}. This argument applies only in the high-energy limit. At lower energies, muon decay and energy loss become important, while comparisons with deep-detector event rates require the corresponding detector response. Moreover, prompt muons can originate not only from charm decays but also from decays of unflavored mesons \cite{Fedynitch2019}. A quantitative constraint on the prompt component therefore requires a spectral and zenith-angle analysis with the relevant production and detector uncertainties \cite{Soldin2019}.

The primary cosmic-ray composition affects atmospheric-muon production through the nucleon flux. In the superposition approximation, the differential nucleon flux at energy $E_N$ is 
\begin{equation}
 \phi_N(E_N)=\sum_A A^2\phi_A(AE_N),
 \label{eq:nucleon_flux}
\end{equation}
where $\phi_A(E_A)$ is the differential flux of nuclei with mass number $A$ per unit total nuclear energy $E_A$, with the same units as $\phi_N$, and $E_A=AE_N$. One factor of $A$ counts the nucleons in a nucleus, while the other comes from $dE_A/dE_N=A$.  For an all-particle power-law spectrum with index $\gamma$ and energy-independent composition fractions $f_A$, the contribution of each nuclear species is weighted by $f_A A^{2-\gamma}$. At $\gamma\simeq2.7$, this becomes $f_A A^{-0.7}$: a heavier composition at fixed all-particle flux reduces the supply of energetic nucleons, even though heavier primaries at fixed total energy generally produce more muons in EAS \cite{Matthews2005}. In reality, the composition fractions vary with energy and must be constrained by measurements \cite{Dembinski2017,Dembinski2025GSF}. Assessing their impact on the predicted muon flux requires propagating the uncertainties of the primary spectrum and composition through the cascade. The possible effect of updated primary-flux parametrizations is also discussed in Ref.~\cite{Zhang2025}.
\bibliography{atm-mu}
\end{document}